\documentclass[twocolumn,showpacs,preprintnumbers,amsmath,amssymb,superscriptaddress,floatfix,nofootinbib]{revtex4}

\usepackage{graphicx}
\usepackage{epsfig}
\usepackage{epstopdf}
\usepackage{hyperref}
\usepackage{amsmath}
\usepackage{amsfonts}
\usepackage{amssymb}

\begin{document}

\title{The $\Omega(2012)$ as a hadronic molecule}
\date{\today}

\author{Ju-Jun Xie}~\email{xiejujun@impcas.ac.cn}
\affiliation{Institute of Modern Physics, Chinese Academy of Sciences, Lanzhou 730000, China}
\affiliation{School of Nuclear Sciences and Technology, University of Chinese Academy of Sciences, Beijing 101408, China}
\affiliation{Southern Center for Nuclear-Science Theory (SCNT), Institute of Modern Physics, Chinese Academy of Sciences, Huizhou 516000, China}

\author{Li-Sheng Geng}~\email{lisheng.geng@buaa.edu.cn}
\affiliation{School of Physics, Beihang University, Beijing 102206, China}
\affiliation{Beijing Key Laboratory of Advanced Nuclear Materials and Physics, Beihang University, Beijing, 102206, China}
\affiliation{Peng Huanwu Collaborative Center for Research and Education, Beihang University, Beijing 100191, China}
\affiliation{Southern Center for Nuclear-Science Theory (SCNT), Institute of Modern Physics, Chinese Academy of Sciences, Huizhou 516000,  China}

\begin{abstract}
	
Recently, a new excited baryon state, $\Omega(2012)$, was observed in the invariant mass spectra of $K^-\Xi^0$ and $K^0_S \Xi^-$ by the Belle Collaboration. This state has a narrow width ($\sim 6$ MeV) compared to other baryon states with a similar mass. In this paper, we  provide a mini-review on the $\Omega(2012)$ state from the molecular perspective, where it appears to be a dynamically generated state with spin-parity $3/2^-$ from the coupled-channels interactions of the $\bar{K} \Xi(1530)$ and $\eta \Omega$ in $s$-wave and $\bar{K} \Xi$ in $d$-wave. Additionally, alternative explanations for the $\Omega(2012)$ resonance are also discussed.

\end{abstract}

\maketitle

\section{Introduction}

In the past decades, hadron physicists showed a great interest in hunting for evidences of exotic states containing some properties that cannot be easily explained by the classical quark models. These exotic states are of special importance in the hadron family. In 2018, a new excited hyperon with strangeness $S=-3$ and isospin $I=0$, the $\Omega(2012)$ state, was discovered by the Belle collaboration in the $K^- \Xi^0$ and $K^0_S \Xi^-$ invariant mass distributions in $\Upsilon(1S)$, $\Upsilon(2S)$, and $\Upsilon(3S)$ decays~\cite{Yelton:2018mag}. Its measured mass and width are $M = 2012.4 \pm 0.7 ({\rm stat}) \pm 0.6  ({\rm syst})$ MeV and $\Gamma = 6.4^{+2.5}_{-2.0}  ({\rm stat}) \pm 1.6  ({\rm syst})$ MeV. The $\Omega(2012)$ is compiled in the \textit{Particle Data Group} (PDG)~\cite{ParticleDataGroup:2022pth} as a three-star excited $\Omega$ state. In addition, the $\Omega(2012)$ has a narrow width and is the first $\Omega$ excited state with preferred negative parity.

\begin{figure}[htbp]
	\centering
	\includegraphics[scale=0.5]{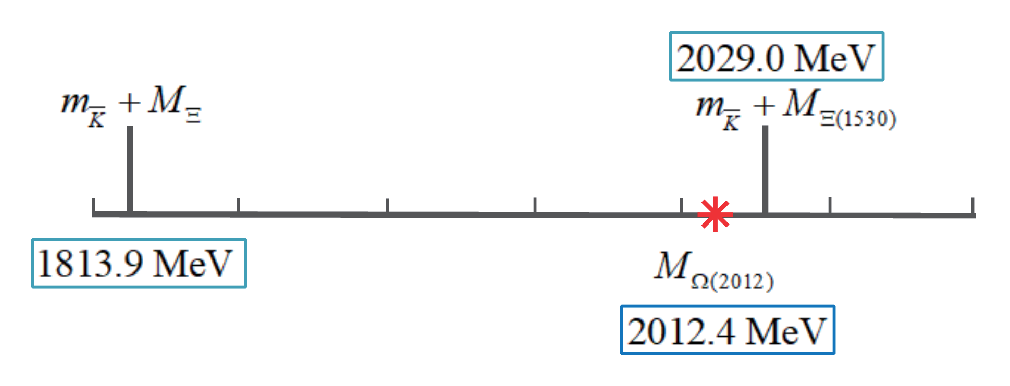}\\
	\caption{The comparision of the experimental mass of $\Omega(2012)$ and the mass thresholds of $\bar{K}\Xi$ and $\bar{K}\Xi(1530)$.}\label{Fig:massrelation}
\end{figure}

The $\Omega$ state contains only three strange quarks, making its excited states challenging to detect. Before the observation of $\Omega(2012)$, most information on the $\Omega$ excited states were extracted from the bubble chamber experiments in the 1980s, and there was only one three-star $\Omega$ excited state, $\Omega(2250)$, with spin-parity ($J^P$) unknown. Meanwhile the other two higher two-star states, $\Omega(2380)$ and $\Omega(2470)$, are very badly known, and they were omitted from the summary table in the PDG~\cite{ParticleDataGroup:2022pth}. The $\Omega(2012)$ resonance has a mass about $340$ MeV higher than the ground state $\Omega(1672)$, and $198.5$ MeV higher than the $\bar{K} \Xi$ mass threshold ($m_{\bar{K}} + M_{\Xi} = 1813.9$ MeV). But, as shown in Fig.~\ref{Fig:massrelation}, the mass of $\Omega(2012)$ is just $16.6$ MeV below the $\bar{K}\Xi(1530)$ mass threshold ($m_{\bar{K}} + M_{\Xi(1530)} = 2029$ MeV), and it is far from the thresholds of $\eta \Omega$ and $\bar{K}\Xi$, which indicates that it could be a possible $\bar{K}\Xi(1530)$ molecule state~\cite{Guo:2017jvc}. In addition, its narrow width implies that its quantum numbers $J^P=3/2^-$ identification is more likely.

In fact, such $\Omega$ excited states have been theoretic studied before the Belle collaboration published their results. For example, using
the chiral unitary approach considering the coupled channels interactions of the $\bar{K}\Xi(1530)$ and $\eta \Omega$ in $s$-wave, the $\Omega$ excited states were investigated~\cite{Kolomeitsev:2003kt,Sarkar:2004jh,GarciaRecio:2006bk,Si-Qi:2016gmh}. A $J^P = 3/2^-$ $\Omega$ excited state with mass around 2012.7 MeV can be dynamically generated~\cite{Si-Qi:2016gmh} with a reasonable value of the subtraction constant: $a_{\mu} = -2.5$ and $\mu = 700$ MeV. However, the obtained width for $\Omega(2012)$ in Ref.~\cite{Si-Qi:2016gmh} is zero, since no open channel (for example, the $\bar{K}\Xi$) was included.

The observation of $\Omega(2012)$ prompted many theoretical investigations to explain its properties with different theoretical models, such as the quark model and the molecular picture on the meson-baryon interaction. For example, its hadronic molecule nature of $\bar{K} \Xi(1530)$ was further investigated in Ref.~\cite{Valderrama:2018bmv} within the one-boson-exchange model and in Refs.~\cite{Lin:2018nqd,Huang:2018wth,Pavao:2018xub,Lin:2019tex,Lu:2020ste} within the chiral unitary approach. It was found that the $\Omega(2012)$ can be dynamically generated from the interactions of $\bar{K} \Xi(1530)$ and $\eta \Omega$ in $s$-wave and $\bar{K}\Xi$ in $d$-wave, and the $\bar{K}\Xi(1530)$ channel is dominant. The measured width of $\Omega(2012)$ can also be reproduced, since the open channel $\bar{K}\Xi$ is also taken into account. However, one should note that the $\eta \Omega$ channel is crucial to produce the $\Omega(2012)$, even its mass threshold ($m_\eta + M_{\Omega} = 2220.3$ MeV) is much higher. Without the $\eta \Omega$ channel, there will be no dynamically generated state because the interaction in the $\bar{K}\Xi(1530)$ and $\eta \Omega$ coupled channels is off diagonal in the chiral unitary approach, which means that, at the leading order, the transition potentials $V_{\bar{K}\Xi(1530) \to \bar{K}\Xi(1530)}$ and $V_{\eta \Omega \to \eta \Omega}$ are zero.

The molecular structure of the $\Omega(2012)$, including five quarks, has its replica in the charm sector. In Ref.~\cite{Ikeno:2023uzz} it is shown that the $\Omega(2012)$ is the analog state of the $\Omega_c(3120)$ discovered by the LHCb Collaboration~\cite{LHCb:2017uwr}.

\section{Three-body decay of $\Omega(2012) \to \bar{K} \Xi^*(1530) \to \bar{K}\pi\Xi$}

\begin{figure}[htbp]
	\centering
	\includegraphics[scale=0.5]{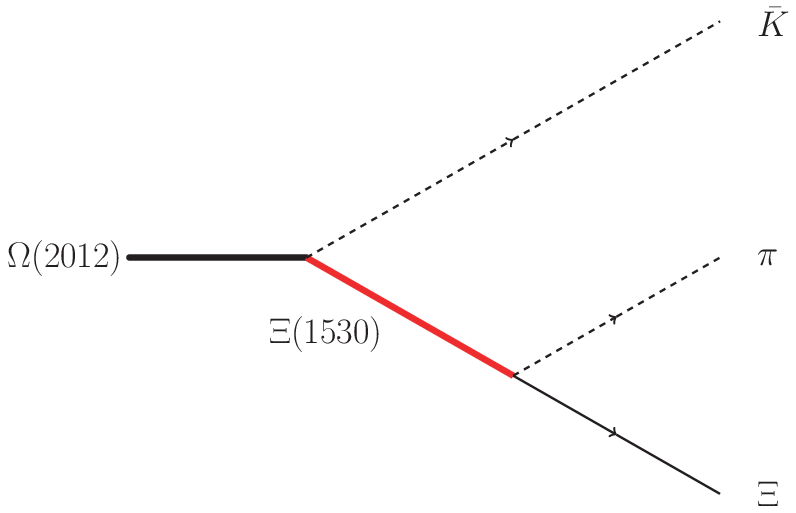}
	\caption{Feynman diagram for the three-body decay of $\Omega(2012) \to \bar{K}\Xi(1530) \to \bar{K}\pi\Xi$.}\label{Fig:three-bodydecay}
\end{figure}

\begin{figure}[htbp]
	\centering
	\includegraphics[scale=0.4]{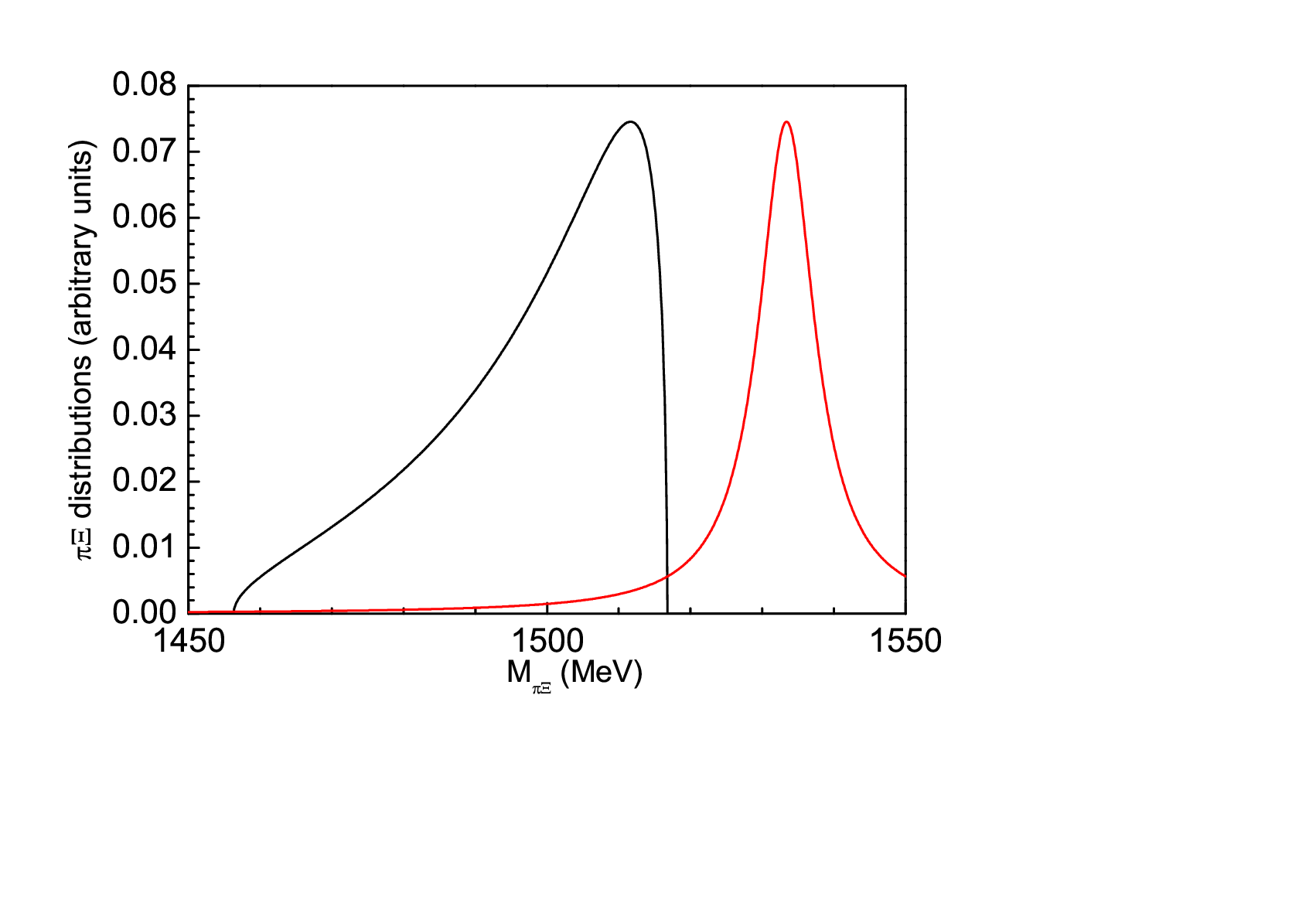} 
	\vspace{-2.5cm}
	\caption{The $\pi\Xi$ invariant mass distributions. The black curve represents the $\pi \Xi$ invariant mass distributions of the $\Omega(2012) \to \bar{K} \Xi(1530) \to \bar{K}\pi \Xi$ decay, while the red curve represents the module squared of $\pi \Xi \to \Xi(1530) \to \pi \Xi$ as a function of the invariant mass $M_{\pi \Xi}$ (see Eq.~\eqref{eq:pixiTsquare}).} \label{Fig:dgdm}
\end{figure}

In the $\bar{K} \Xi(1530)$ molecular picture, as shown in Fig.~\ref{Fig:three-bodydecay}, the three-body decay mode of $\Omega(2012) \to \bar{K} \Xi(1530) \to \bar{K}\pi\Xi$ is important, and the three-body decay width for $\Omega(2012) \to \bar{K}\pi\Xi$ is comparable with the one for $\Omega(2012) \to \bar{K} \Xi$. The $\pi \Xi$ invariant mass distributions can be written as

\begin{eqnarray}
	\frac{d\Gamma_{\Omega(2012) \to \bar{K} \Xi(1530) \to \bar{K} \pi\Xi}}{dM_{\pi\Xi}} \propto  \frac{1}{M} \frac{ p_{\bar K} p^*_\pi \Gamma_{\Xi^*} }{(M_{\pi\Xi} - M_{\Xi^*})^2 + \Gamma_{\Xi^*}^2/4},
\end{eqnarray}
where $M_{\Xi^*}$ and $\Gamma_{\Xi^*}$ are the mass and width of $\Xi(1530)$, and $M_{\pi\Xi}$ is the invariant mass of the $\pi$ and $\Xi$ system. While $p_{\bar{K}}$ and $p^*_\pi$ are the three-momentum of $\bar{K}$ and $\pi$ in the $\Omega(2012)$ rest frame and the $\pi$ and $\Xi$ center of mass frame, respectively. Then the invariant $\pi \Xi$ mass distributions can be easily obtained as shown by the black curve in Fig.~\ref{Fig:dgdm}. The numerical results are calculated with averaged values for the masses of those involved particles, and we take $\Gamma_{\Xi^*} = 9.5$ MeV.

In the molecular picture, the production of the $\Omega(2012)$ resonance in the nonleptonic weak decays of $\Omega_c^0 \to \pi^+ \bar{K}\Xi(1530) (\eta \Omega) \to \pi^+ (\bar{K}\Xi)^-$ and $\pi^+ (\bar{K}\Xi\pi)^-$ via final-state interactions of the $\bar{K}\Xi(1530)$ and $\eta \Omega$ pairs was investigated~\cite{Zeng:2020och,Ikeno:2022jpe}. Later on, the theoretical results of Ref.~\cite{Zeng:2020och} were confirmed by the Belle Collaboration~\cite{Belle:2021gtf}. The significance of the $\Omega(2012)$ signal in the $\Omega^0_c \to \pi^+ \Omega^-(2012) \to \pi^+ (\bar{K}\Xi)^-$ is $4.2\sigma$, including the systematic uncertainties. Further experimental studies about this decay channel are most welcome.

On the other hand, because of the small phase space [the mass of $\Omega(2012)$ is below the mass threshold of $\bar{K}\Xi(1530)$ and the small decay width ($\sim 10$ MeV) of $\Xi(1530)$], it is difficult to analyze the three-body decay of $\Omega(2012) \to \bar{K} \Xi(1530) \to \bar{K}\pi\Xi$. To see this, we define
\begin{eqnarray}\label{width3b}
\frac{d|T|^2_{\pi \Xi \to \Xi(1530) \to \pi\Xi}}{dM_{\pi\Xi}} \propto  \frac{ M_{\Xi^*} \Gamma_{\Xi^*} }{(M_{\pi\Xi} - M_{\Xi^*})^2 + \Gamma_{\Xi^*}^2/4}. \label{eq:pixiTsquare}
\end{eqnarray}
The obtained $\pi \Xi$ invariant mass distribution is also shown in Fig.~\ref{Fig:dgdm} by the red curve. One can clearly see that the two peaks (we have normalized them to the same value) are much separated and the overlapping part of the two line shapes is very narrow, then the $\Omega(2012) \to \bar{K}\Xi(1530) \to \bar{K}\pi\Xi$ decay will be suppressed due to the highly off-shell effect of the $\Xi(1530)$ propagator, though the coupling of $\Omega(2012)$ to the $\bar{K}\Xi(1530)$ channel is strong~\cite{Lu:2020ste,Arifi:2022ntc}. Therefore, in the real analysis, the physical masses for the $\Xi(1530)$, $\bar{K}$, $\pi$, and $\Xi$ with different charges should be taken into account explicitly~\cite{Huang:2018wth}.  

It is worth mentioning that in an early search for the $\Omega(2012) \to \bar{K}\Xi(1530) \to \bar{K} \pi\Xi$ decay by the Belle collaboration~\cite{Belle:2019zco}, it did not find the signal and set an upper limit at $90\%$ confidence level on the branching fraction ratio
\begin{equation}
{\cal R}^{\bar{K}\pi\Xi}_{\bar{K}\Xi}  =  \frac{ {\rm Br}[\Omega(2012) \to \bar{K}\Xi(1530) \to \bar{K}\pi\Xi]}{{\rm Br}[\Omega(2012) \to \bar{K}\Xi]} < 11.9\% .
\end{equation} 

While, in the very recent measurement of the Belle Collaboration~\cite{Belle:2022mrg}, the ratio ${\cal R}^{\bar{K}\pi\Xi}_{\bar{K}\Xi}$ was updated with improved selection criteria, accounting for the $\Omega(2012)$ being below the $\bar{K}\Xi(1530)$ mass threshold. In this new analysis, $M_{\pi \Xi} < 1.517$ GeV is required and the signal shape of $\Omega(2012)$ was parameterized with a Flatt\'e-like function to account for the allowed phase space. Such selection criteria were not included in the previous analysis in Ref.~\cite{Belle:2019zco}, and they improve the signal-background sepration and hence increase the signal yield. The new result of the ratio ${\cal R}^{\bar{K}\pi\Xi}_{\bar{K}\Xi}$ is $0.97 \pm 0.24 \pm 0.07$. This is consistent with the molecular interpretation for the $\Omega(2012)$ as proposed in Refs.~\cite{Lin:2018nqd,Huang:2018wth,Pavao:2018xub,Lin:2019tex,Lu:2020ste}. In the molecular picture for $\Omega(2012)$, there are two free model parameters for the $d$-wave interaction. By adjusting these two model parameters, it is found that the ratio ${\cal R}^{\bar{K}\pi\Xi}_{\bar{K}\Xi}$ reported recently by the Belle Collaboration can be also easily accommodated.

\section{Two-body decay of $\Omega(2012) \to \bar{K}\Xi$}

\begin{figure}[htbp]
	\centering
	\includegraphics[scale=0.8]{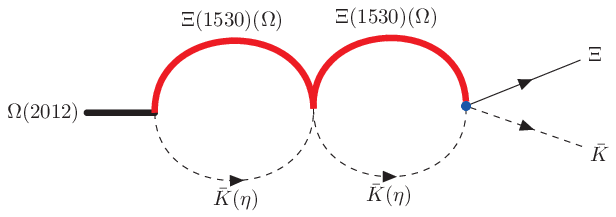}\\
	\caption{Feynman diagram for the effective two-body decay of $\Omega(2012) \to \bar{K}\Xi$ through the re-scattering of $\bar{K}\Xi(1530)$ and $\eta \Omega$ in coupled channels.}\label{Fig:two-body-bubble}
\end{figure}

In Ref.~\cite{Lu:2020ste}, the $\Omega(2012)$ state was investigated from the molecular perspective in which the resonance is dynamically generated from the interactions of $\bar{K}\Xi^*(1530)$, $\eta \Omega$ and $\bar{K}\Xi$ in coupled channels, where the interactions of $\bar{K}\Xi^*(1530)$ and $\eta \Omega$ are in $s$-wave, and the $\bar{K}\Xi$ interaction is in $d$-wave. One has to include further parameters for building the transition potentials of $V_{\bar{K}\Xi(1530) \to \bar{K}\Xi}$ and $V_{\eta \Omega \to \bar{K}\Xi}$. Fortunately, these parameters can be determined from the measured mass and width of the $\Omega(2012)$ state. Furthermore, for the effective $\Omega(2012) \bar{K} \Xi$ vertex, it can be obtained from the $s$-wave $\Omega(2012)\bar{K}\Xi^*(1530)$ and $\Omega(2012)\eta\Omega$ interactions and the re-scattering of $\bar{K}\Xi^*(1530)$ and $\eta\Omega$ pairs, which proceed as shown in Fig.~\ref{Fig:two-body-bubble}. In this respect, the final state interactions of $\bar{K}\Xi \to \bar{K}\Xi$ is already taken into account in the effective coupling of the $\Omega(2012)\bar{K}\Xi$ vertex.

At the pole position of $\Omega(2012)$, one can extract the effective strong coupling of $\Omega(2012)$ to the $\bar{K}\Xi$ channel, $g_{\Omega(2012)\bar{K}\Xi}$, from the unitarized scattering amplitude $T_{\bar{K}\Xi \to \bar{K}\Xi}$, which was obtained by solving the Bethe-Salpeter equation and considering the re-scattering of all the three coupled channels. Close to the pole at $z_R = M_R - i \Gamma_R/2$, the scattering amplitude $T_{\bar{K}\Xi \to \bar{K}\Xi}$ can be written as
\begin{eqnarray}
T_{\bar{K}\Xi \to \bar{K}\Xi} = \frac{|g_{\Omega(2012)\bar{K}\Xi}|^2}{\sqrt{s} - z_R}. 
\end{eqnarray}
Then the partial decay width of the $\Omega(2012) \to \bar{K}\Xi$ can be easily obtained with
\begin{equation}\label{width2b}
	\Gamma_{\Omega(2012) \rightarrow \bar{K}\Xi} =
	\frac{|g_{\Omega(2012)\bar{K}\Xi}|^2}{2\pi} \frac{M_{\Xi}}{M}q_{\bar{K}},
\end{equation}
where $g_{\Omega(2012)\bar{K}\Xi}$ is the effective coupling constant of the $\Omega(2012)\bar{K}\Xi$ vertex obtained as explained above, and $q_{\bar{K}}$ is the momentum of $\bar{K}$ in the rest frame of $\Omega(2012)$, which is
\begin{eqnarray}
	q_{\bar{K}} = \frac{\sqrt{[M^2 -(m_{\bar{K}} + M_{\Xi})^2][M^2 -
			(m_{\bar{K}} - M_{\Xi})^2]}}{2M} .
\end{eqnarray}

In general, since the $\Omega(2012) \to \bar{K}\Xi$ decay is in $d$-wave, the $\Gamma_{\Omega(2012) \rightarrow \bar{K}\Xi}$ should be proportional to $q^5_{\bar{K}}$, but the transition amplitudes $T_{\bar{K}\Xi(1530) \to \bar{K}\Xi}$ and $T_{\eta \Omega \to \bar{K}\Xi}$ potentials incorporate the four extra powers of $q_{\bar{K}}$. (More details can be found in Ref.~\cite{Pavao:2018xub} and in Refs.~\cite{Sarkar:2005ap,Roca:2006sz} for the case of the $\Lambda(1520)$ resonance where the interactions of $\bar{K}N$ and $\pi\Sigma$ in $d$-wave are included.) In this way, the obtained effective coupling constant $g_{\Omega(2012)\bar{K}\Xi}$ is dimensionless. Assuming that only the three body and two-body strong decays contribute to the total decay width of $\Omega(2012)$, and with the new measurements for the ratio ${\cal R}^{\bar{K}\pi\Xi}_{\bar{K}\Xi}$, we obtain
\begin{eqnarray}
|g_{\Omega(2012)\bar{K}\Xi}| = 0.28 \pm 0.14 , 
\end{eqnarray}
with the error obtained from the uncertainties of ${\cal R}^{\bar{K}\pi\Xi}_{\bar{K}\Xi}$ and the total decay width of $\Omega(2012)$.

\section{Other explanations of the $\Omega(2012)$}

On the other hand, although the $\Omega(2012)$ resonance was proposed to be a dynamically generated state, it is difficult to be clearly distinguished from the classical quark model state due to the possible large mixing of molecular component and the $sss$ configuration. In fact, the $\Omega$ excited states were also investigated in classical quark models~\cite{Capstick:1986bm,Loring:2001ky,Pervin:2007wa,Chen:2009de,Faustov:2015eba} and in the five-quark picture~\cite{Yuan:2012zs,An:2013zoa,An:2014lga}. In a pioneering study~\cite{Chao:1980em}, the $\Omega$ excited states were predicted employing a quark model with ingredients suggested by QCD, and the mass of one predicted state with $J^P=3/2^-$ is about $2020$ MeV.

In the framework of the chiral quark model, the authors of Refs.~\cite{Liu:2020yen,Hu:2022pae} performed dynamical calculations of pentaquark systems with $sssu\bar{u}$ and $sss \bar{d}d$, and conclude that the $\Omega(2012)$ can be interpreted as a $\bar{K}\Xi(1530)$ molecular state with isospin zero and spin-parity $J^P = 3/2^-$. The mass and two-body strong decays of the $\Omega(2012)$ were also studied within the quantum chromodynamics sum rules~\cite{Aliev:2018yjo,Aliev:2018syi,Su:2024lzy} and constituent quark model~\cite{Xiao:2018pwe,Wang:2018hmi,Liu:2019wdr,Wang:2022zja,Zhong:2022cjx}, and these studies showed that the $\Omega(2012)$ could be a good candidate of the first orbital $(1P)$ excitations of the ground-state $\Omega$ baryon with $J^P=3/2^-$. A spin-orbit partner with $J^P = 1/2^-$ of $\Omega(2012)$ was naturally predicted in Ref.~\cite{Arifi:2022ntc} within the constituent quark model. In addition, within a nonretivistic constituent quark potential model, a $J^P= 1/2^-$ state $\Omega(1^2P_{1/2^-})$ with mass around 1950 MeV and width about 12 MeV was also predicted~\cite{Liu:2019wdr}. Such a $J^P=1/2^-$ state is hard to be generated in the molecular scenario of $\Omega(2012)$. Therefore, the discovery of the spin-orbit partner with $J^P = 1/2^-$ would be crucial in distinguishing between the three-quark model and the hadronic molecular nature for the $\Omega(2012)$ state.
 
Meanwhile, in Ref.~\cite{Lu:2022puv}, a systematic analysis of the mass and strong decays of the $\Omega(2012)$ resonance was performed in a coupled-channel approach of a bare three-quark state and a composite meson-baryon state. Besides the coupling of the bare state and $\bar{K}\Xi(1530)$ channel, the effective meson-baryon interactions are also included. It was found that both the three-quark core and $\bar{K}\Xi(1530)$ channel are essential for the description of $\Omega(2012)$ resonance.

\section{Propects for the obeservation of $\Omega(2012)$ from its radiative decay of $\Omega(2012) \to \gamma \Omega$}

\begin{figure}[htbp]
	\centering
	\includegraphics[scale=1.2]{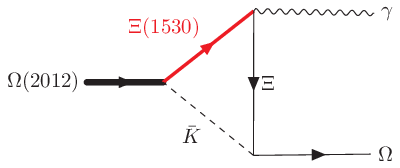}\\
	\caption{The radiative decay of $\Omega(2012) \to \gamma \Omega$ through the triangle diagram.}\label{Fig:radiative}
\end{figure}

Within the three quark model, in Ref.~\cite{Liu:2019wdr}, it was found that the $\Omega(2012)$ resonance is most likely to be a $J^P=3/2^-$ 1P-wave state $\Omega(1^2P_{3/2^-})$, and it was pointed out that the $\Omega(2012)$ has a large potential to be observed in the $\gamma \Omega$ channel. While, within the molecular picture, the ratiative decay of $\Omega(2012) \to \gamma \Omega$ can also be proceed with the triangle diagram as shown in Fig.~\ref{Fig:radiative}. The radiative decay of $\Omega(2012)$ is difficult to be measured by the Belle Collaboration, but, it could be done by the BESIII Collaboration and the planed high-luminosity electron-positron collision machine Super $\tau$-Charm Facility~\cite{Guo:2022kdi}.

Until now, only the mass and width of the $\Omega(2012)$ resonance were measured, more experimental information is really needed. For example, the angular distributions for its two-body strong decay of $\Omega(2012) \to \bar{K}\Xi$, from where its $J^P$ quantum numbers can be determined. Furthermore, both on theoretical and experimental sides, such studies on the radiative decay of $\Omega(2012) \to \gamma \Omega$ are mostly welcome, and they will be help to further understand the nature of the $\Omega(2012)$ state. 

\section{Summary}

The discovery of $\Omega(2012)$ state has attracted many theoretical works, trying to understand its nature with different models. In the molecular picture, the $\Omega(2012)$ state appears to be dynamically generated from the coupled channels interactions of the $\bar{K} \Xi(1530)$ and $\eta \Omega$ in $s$-wave and $\bar{K} \Xi$ in $d$-wave. In such a scenario, the $\Omega(2012)$ is interpreted as a $3/2^-$ molecule state, and most of its properties can be well accommodated. More precise experimental measurements on the strong and radiative decays of the $\Omega(2012)$ state are expected to provide valuable insights into its nature. These measurements can be carried out not only by the Belle collaboration but also by the Belle II, LHCb, and BESIII Collaborations in the near future, as well as by the next generation facilities~\cite{KLF:2020gai,Aoki:2021cqa}. Searching for its symmetry and three-body counterparts as well as studying the $\eta \Omega$ and $\bar{K}\Xi(1530)$ interactions using the femtoscopic technique will also be useful in verifying the molecular nature of $\Omega(2012)$~\cite{Liu:2024uxn}.

\section*{Acknowledgments}

We would like to thank Profs. Eulogio Oset, Bing-Song Zou, and Cheng-Ping Shen for fruitful discussions. This work is partly supported by the National Key R\&D Program of China under Grant No. 2023YFA1606703 and by the National Natural Science Foundation of China under Grant Nos. 12075288 and 12361141819. It is also supported by the Youth Innovation Promotion Association CAS.

\end{document}